# Comparative CFD modelling of the interplay between aqueous humour hydrodynamics and drug and nanocarrier transport for glaucoma drug delivery via intracameral injection, drug-eluting implants and contact lenses

Nazifa Begum[1], Robert Diniţă[1,†], Naman Patel[1,†], Umar Pitafi[1], Tao Chen[1], Cynthia Yu-Wai-Man[2,*], Daniel Sebastia-Saez[1,*]

[1]School of Chemistry and Chemical Engineering, University of Surrey, Guildford, GU2 7XH, UK

[2]Faculty of Life Sciences & Medicine, King's College London, London, SE1 1UL, UK

[†]These authors contributed equally to this work

*Corresponding Authors:

Daniel Sebastia-Saez, University of Surrey, j.sebastiasaez@surrey.ac.uk

Cynthia Yu-Wai-Man, King's College London, cynthia.yu-wai-man@kcl.ac.uk

**List of symbols**

*Latin symbols*

$A$ – Area [m$^2$]

$c$ – Concentration [mol/m$^3$]

$c_p$ – Specific heat capacity [J/kg/K]

$g$ – Gravity acceleration [m/s$^2$]

$J$ – Mas flux [mol/m$^2$/s]

$k$ – Thermal conductivity [W/m/K]

$M_{dose}$ – Dose [g]

$M_w$ – Molecular weight [g/mol]

$p$ – Pressure [Pa]

$\boldsymbol{r}$ – Position vector [m]

$T$ – Temperature [K]

$t$ – Time [s]

$\boldsymbol{u}$ – Velocity [m/s]

*Greek symbols*

$\varepsilon$ – Emissivity (dimensionless)

$\gamma$ – Drug release rate from implant [1/s]

$\beta$ – Coefficient of thermal expansion [1/K]

$\mu$ – Dynamic viscosity [Pa s]

$\theta$ – Saccade angle [rad]

$\sigma$ – Boltzmann constant [$1.380649\times10^{-23}$ J/K]

$\rho$ – Density [$kg/m^3$]

$\tau$ – Release timescale [s]

$\omega$ – Angular velocity [rad/s]

**List of abbreviations**

CFD – Computational Fluid Dynamics

CV – Coefficient of Variation

FDA – Food and Drug Administration

IOP – Intraocular Pressure

PC – Posterior chamber

PLGA – poly(lactic-*co*-glycolic) acid

SC – Schlemm's Canal

TM – Trabecular meshwork

## Abstract

Glaucoma is the leading cause of irreversible blindness worldwide and often requires long-term administration of intraocular pressure-lowering agents. The effectiveness of ocular drug delivery depends not only on the delivery route, but also on the transport mechanisms governing drug distribution within the eye. In this study, a computational fluid dynamics (CFD) framework was developed to investigate the interplay between aqueous humour hydrodynamics and the transport of both dissolved drug and nanocarriers delivered through three mechanisms, including intracameral injection, drug-eluting implants and contact-lens administration. An idealised three-dimensional model of the anterior segment was implemented in COMSOL Multiphysics v6.4, incorporating aqueous humour flow, thermal convective effects, drug transport, saccadic movements and particle tracing. This approach enabled comparison between molecular diffusion-dominated and particle-mediated delivery. The simulations revealed substantial differences in intraocular transport behaviour among the delivery strategies. Contact-lens delivery produced the most spatially homogeneous distributions, whereas implant-based delivery generated persistent concentration gradients associated with localised release. Intracameral injection exhibited intermediate behaviour, with rapid redistribution of the administered agent by aqueous humour circulation. Nanocarrier transport exhibited greater spatial heterogeneity within the anterior chamber than dissolved-drug transport owing to the reduced particle diffusivity and the resulting advection-dominated transport. The results demonstrate that aqueous humour hydrodynamics constitute an independent determinant of ocular drug transport. Accounting for these physiological effects may therefore improve the design and optimisation of sustained ophthalmic drug delivery systems. The proposed modelling framework will provide a useful tool for the design and optimisation of diluted-drug and nanocarrier-based glaucoma drug delivery systems in the future.

## 1 Introduction

Glaucoma is a progressive optic neuropathy and the leading cause of irreversible blindness worldwide, affecting approximately 95 million people[1]. The disease is characterised by the gradual degeneration of the optic nerve, resulting in visual field loss. Elevated intraocular pressure (IOP) remains the most significant risk factor for glaucoma progression. This occurs due to obstruction of the aqueous humour outflow pathway, and current therapeutic strategies are primarily directed towards reducing IOP through pharmacological or surgical interventions. Topical ophthalmic medications, particularly prostaglandin analogues like bimatoprost, are widely prescribed treatments due to their effectiveness in enhancing aqueous humour outflow and lowering IOP[2]. However, despite their clinical efficacy, conventional eye-drop formulations exhibit significantly poor ocular bioavailability, with less than 5% of the administered drug typically penetrating into intraocular tissues[3]. This inefficiency arises from multiple physiological barriers, including tear turnover, blinking, nasolacrimal drainage and limited corneal permeability, resulting in rapid drug elimination from the ocular surface[4,5].

In addition to poor bioavailability, topical therapies depend heavily on patient adherence, which is a major challenge in glaucoma management. Since glaucoma is generally asymptomatic in its early stages, many patients fail to maintain consistent dosing regimens, leading to inadequate IOP control and progressive visual deterioration[6,7]. Consequently, sustained and targeted ocular drug delivery systems have emerged as an important area of research, aiming to improve the therapeutic efficacy while reducing the administration frequency and improving patient compliance.

Among the most promising alternatives to topical administration are intracameral injections, biodegradable implants and drug-eluting contact lenses[8–10]. Intracameral injections enable direct delivery of therapeutics into the anterior chamber, bypassing many of the precorneal barriers associated with topical administration. However, injected drug molecules are still subjected to aqueous humour turnover and trabecular meshwork outflow, which often results in rapid clearance and short therapeutic duration[11]. To address these limitations, sustained-release intracameral implants have been developed, offering prolonged drug release directly within the eye. A clinically significant advancement is the biodegradable intracameral implant Durysta®, containing bimatoprost, which was approved by the United States Food and Drug Administration (FDA) for the treatment of open-angle glaucoma[2]. Such implants exemplify the growing clinical relevance of sustained ocular delivery technologies.

Another increasingly investigated platform involves therapeutic contact lenses capable of sustained drug release across the corneal surface. These drug-eluting contact lenses extend the residence time of the drug on the ocular surface compared with conventional eye drops, potentially increasing drug absorption and reducing dosing frequency[12]. Additionally, the incorporation of nanocarriers into contact lenses has been shown to modulate release kinetics, improve drug stability and enhance ocular penetration[4,13]. Nanocarrier systems enable the encapsulation of anti-glaucoma agents, supporting controlled release, enhanced ocular penetration and reduced systemic exposure[14].

Despite substantial experimental progress in sustained ocular drug delivery, direct comparisons between different administration strategies remain difficult *in vivo* due to the complexity of ocular physiology and the challenges of measuring intraocular transport dynamics experimentally. In this context, Computational Fluid Dynamics (CFD) modelling has emerged as a powerful tool for investigating ocular transport phenomena. CFD enables the simulation of aqueous humour hydrodynamics, drug diffusion and particle transport within anatomically representative ocular geometries, while allowing systematic evaluation of different delivery strategies under controlled conditions.

Several previous studies have employed CFD approaches to investigate ocular fluid flow and drug transport. Investigations of aqueous humour dynamics within the anterior chamber have demonstrated the importance of buoyancy-driven flow, trabecular meshwork outflow and thermal gradients in determining intraocular transport behaviour. Early numerical studies utilised the Navier-Stokes equations to characterise aqueous humour circulation and pressure[15]. Findings revealed recirculatory laminar flow patterns and the influence of buoyancy-driven convection, both of which can affect drug residence time and particle trajectories within the anterior chamber. Subsequent research introduced anatomically realistic three-dimensional CFD models; for example, sections of the human eye using various imaging techniques have been reconstructed to analyse velocity fields and IOP distribution under glaucomatous conditions[16]. CFD models have been useful in unveiling that the geometry of the outflow pathway significantly affects local shear stresses and pressure distribution[17]. While at a smaller scale, it has also been found that high-resolution fluid-structure interaction (FSI) models have elucidated the roles of the trabecular meshwork (TM) and Schlemm's canal (SC) in regulating flow and mechanical stresses[18].

Parallel developments have occurred in nanoparticle transport modelling. Eulerian–Lagrangian approaches combining continuum fluid flow with discrete particle tracing have been utilised to investigate nanoparticle trajectories, residence times and deposition patterns in biological systems. However, this has yet to be extensively explored in ocular drug delivery applications. In ocular applications, particle-tracing methods can enable evaluation of how nanocarrier size, buoyancy and fluid recirculation affect transport pathways through the anterior chamber. Such methods provide important mechanistic insight into the behaviour of nanocarrier-based glaucoma therapies, particularly when comparing different routes of administration.

Although previous computational studies have examined individual ocular drug delivery systems, there remains a lack of comprehensive and comparative analyses evaluating drug diffusion and nanocarrier transport across multiple clinically relevant administration routes within a unified computational framework[19]. In particular, limited work has directly compared intracameral injection, intracameral implant and contact-lens-mediated delivery under consistent physiological and hydrodynamic conditions. Furthermore, few studies have simultaneously examined both dissolved drug transport and nanocarrier trajectories, despite the increasing prominence of nanoparticle-enabled ophthalmic therapeutics. As a result, the relative influence of aqueous humour hydrodynamics on different delivery strategies remains insufficiently understood.

Therefore, the present study develops a comparative CFD framework to investigate glaucoma drug delivery via intracameral injection, drug-eluting implant and contact-lens administration using COMSOL Multiphysics 6.4. Both dissolved bimatoprost transport and nanocarrier-based delivery are considered, enabling direct comparison between conventional molecular transport and particle-mediated transport within a common anatomical and physiological environment. An idealised three-dimensional model of the anterior segment is implemented, incorporating aqueous humour flow, thermal buoyancy effects, solute transport and Lagrangian particle tracing.

To isolate the influence of aqueous humour hydrodynamics on transport behaviour, equivalent total drug doses were considered for all delivery strategies. The study evaluates the effect of administration route on spatial drug distribution, spatial homogeneity and intraocular retention within the aqueous humour. For the nanocarrier systems, particle trajectories are additionally analysed to elucidate the role of advection-dominated transport. By comparing dissolved drug and nanoparticle transport under identical physiological conditions, the proposed framework provides mechanistic insight into the interplay between delivery strategy and aqueous humour hydrodynamics, thereby supporting the improvement of future sustained-release and nanocarrier-enabled glaucoma therapies.

The present study investigates the role of aqueous humour hydrodynamics as an independent determinant of ocular drug transport by comparing three representative glaucoma drug delivery strategies, namely intracameral injection, drug-eluting implant positioned at the iridocorneal angle, and contact lens delivery. Dissolved drug transport and nanocarrier transport are examined separately to distinguish the influence of molecular diffusion from advection-dominated particle transport. Rather than assessing formulation-specific performance, the objective is to isolate and quantify how physiological aqueous humour flow governs drug retention, spatial distribution and transport homogeneity within the anterior segment. This mechanistic understanding may assist the rational design and optimisation of future ocular drug delivery systems.

## 2 Methodology

### 2.1 Ocular Geometry and Computational Domain

An idealised three-dimensional computational model of the human eye anterior segment was developed using COMSOL Multiphysics v6.4. The geometry was constructed with built-in modelling tools and anatomical parameters obtained from the literature, as summarised in Table 1.

*Table 1 Geometry parameters of the anterior segment*

| Parameter | Value [mm] |
| --- | --- |
| Radius of posterior cornea [20] | 6.80 |
| Maximum height of anterior segment [20] | 2.63 |
| Lens radius [20] | 10.00 |
| Height of iris-lens channel[20–22] | 0.093 |
| Iris thickness [23] | 0.40 |
| Pupil radius [22] | 1.75 |

The computational domain included the anterior chamber (AC), extending from the corneal endothelium to the iris and the posterior chamber (PC), located between the iris and the lens. The trabecular meshwork (TM) was modelled as a porous region to simulate aqueous humour outflow, while the ciliary body region was defined as the inflow boundary for aqueous humour secretion (Figure 1). Gravity was assumed constant and aligned with the vertical axis (x-axis) corresponding to the upright orientation of the eye to obtain the results in this work, except for the validation graph where a case was run in the supine position.

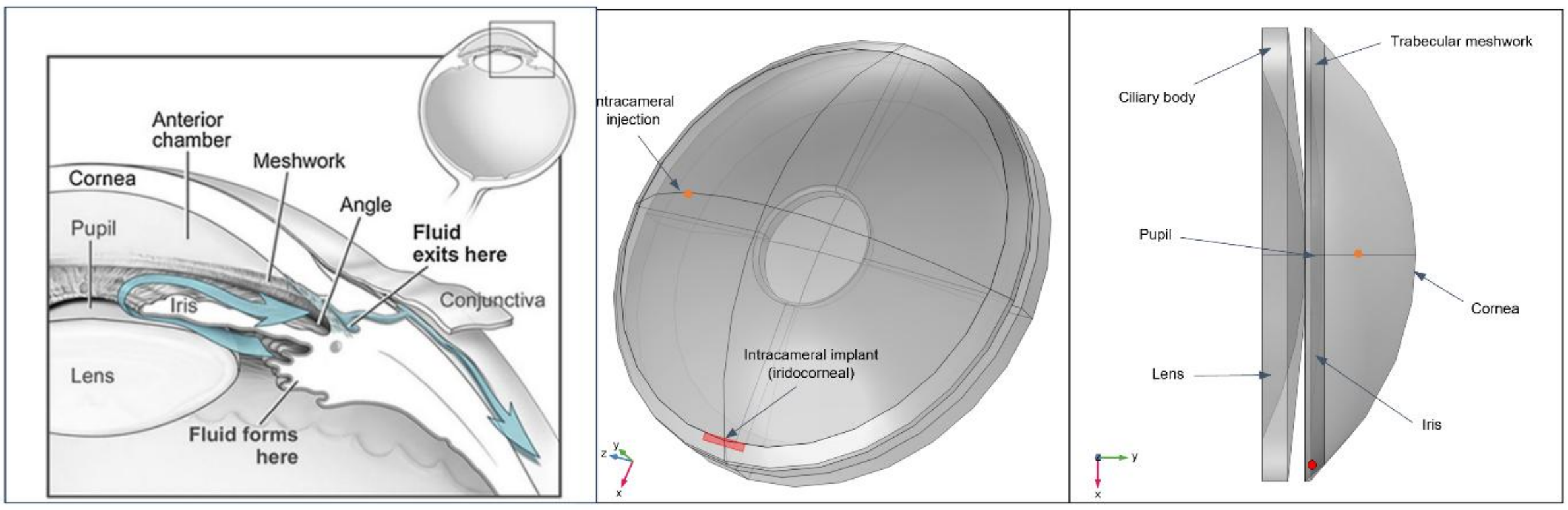

*Figure 1 Schematic illustration of the calculation domain.*

In the implant-based delivery scenario, an intracameral implant geometry was incorporated to represent sustained drug-delivery devices. The dimensions of Durysta® were implemented as it is the only FDA-approved biodegradable intracameral implant[24]. For the injection-based scenario, a localised point source within the anterior chamber was defined for the release of both the drug and nanocarriers. The injection location was defined based on intracameral administration guidance provided for Durysta®[24]. In the case of contact lens delivery, a flux boundary was imposed on the cornea such that the mass of drug introduced into the anterior chamber during delivery time equalled the dose in the other two delivery methods tested (i.e. equalled the initial dose in the implant and the dose injected). This approach allows isolating the effect of aqueous humour hydrodynamics from complex transcorneal transport.

## 2.2 Governing Equations

### Aqueous Humour Flow

Static simulations were run to obtain the velocity field prior to solving the transient effects of drug delivery. Aqueous humour flow was modelled as an incompressible, Newtonian single phase laminar flow. The flow field was computed using the laminar flow physics interface in COMSOL Multiphysics v6.4. The conservation of momentum equation is:

$$\rho(\boldsymbol{u} \cdot \nabla)\boldsymbol{u} = -\nabla \mathrm{p} + \mu\nabla^2 u + \rho\boldsymbol{g}\left[1 - \beta(T - T_{ref})\right] \tag{1}$$

And the conservation of mass:

$$\rho\nabla \cdot \boldsymbol{u} = 0 \tag{2}$$

Where $\rho$ represents the density, $\boldsymbol{u}$ is the velocity vector, $p$ is the pressure, $\mu$ is the dynamic viscosity and $\boldsymbol{g}$ is the acceleration due to gravity. The thermally driven buoyant force was defined using the Boussinesq approximation. The trabecular meshwork and the implant were represented as porous subdomains, where the local momentum balance was modified according to Darcy's law. The implant was assigned a porosity of 0.9 (i.e., PLGA scaffold)[25]. The geometric and flow parameters of the trabecular meshwork were obtained from the available literature[26].

A no-slip boundary condition was implemented in the rest of the surfaces, including the cornea, lens and iris. An outlet boundary condition was set on the trabecular meshwork. A uniform inflow velocity of $v_0$ = 2.03 μm·s$^{-1}$ was imposed at the aqueous humour inlet, corresponding to a physiological AH production rate of approximately 2.5 μl/min[27]. A static intraocular pressure of 27 mmHg was applied to represent glaucomatous conditions[20].

**Heat Transfer**

The heat transfer formulation employed in this study can be directly related to Pennes' Bioheat equation:

$$\nabla \cdot (k \nabla T) = 0 \tag{3}$$

Where the terms that account for blood perfusion and metabolic heat generation within biological tissues have been neglected, since in the case of aqueous humour, these effects are negligible because the fluid is avascular and has no metabolic activity. Furthermore, the simulations were conducted under static conditions allowing the transient term to be omitted. The symbol $k$ denotes the thermal conductivity of aqueous humour. Imposed boundary conditions were constant temperatures in all surfaces (37°C at the inner surface of the cornea and 34°C at the rest of the surfaces including lens and iris).

**Drug Transport**

The steady-state thermal convective flow was used as the initial condition to solve the transient transport equation in the second step of the simulations using the equation:

$$\frac{\partial c}{\partial t} + \nabla \cdot \boldsymbol{J} + \boldsymbol{u} \cdot \nabla c = R \tag{4}$$

Where $c$ is the concentration of the drug and $\boldsymbol{J} = -D\nabla c$ accounts for Fick's diffusion flux. The source term $R$ accounts for the delivery of the drug from the implant and into the bulk fluid flow of aqueous humour. The diffusion coefficient in water was calculated using the well-known Stokes-Einstein equation.

Where $T$ = 34.51°C is the temperature of the eye and $\mu$ = 0.7269 mPa·s is the dynamic viscosity of water at that temperature. The parameter $k_b$=1.380649×10$^{-23}$ J/K is the Boltzmann constant. The molecular radius $r$ was calculated using the empirical expression:

Which assumes spherical symmetry of the bimatoprost molecule, used across methodologies tested.

**Delivered dose**

To facilitate direct comparison between the three delivery strategies, the total administered drug $M_{dose} = 10$ μg was held constant across all simulations and delivery methods. Differences in predicted drug distribution thus arose from the delivery route and subsequent aqueous humour transport, rather than from differences in the administered dose or losses due to tear production, blinking and complex corneal barrier effects, which were excluded from this study. Different delivery routes were implemented through distinct initial and boundary conditions. Intracameral injection was represented as a localised initial concentration region within the anterior chamber. Implant delivery was modelled as diffusion-driven release from a porous implant domain. For implant-based delivery, bimatoprost presents a low solubility in the aqueous humour, therefore, it is not readily available in the implant for dissolution. Instead, a hydrolysis process takes place before the drug can be carried away by the aqueous humour. Delivery from the implant was implemented using the following first-order equation, where an umbrella parameter $\beta$ has been defined to represent chemical phenomena leading to bimatoprost dissolution:

$$\frac{\partial c_{un}}{\partial t} = -\gamma c_{un} \tag{5}$$

Where $c_{un}$ is the concentration of undissolved drug in the implant. Since the objective of the study was to evaluate the influence of aqueous humour hydrodynamics on drug transport rather than to optimise the implant release kinetics, the results were analysed using time normalised by the characteristic release duration as well as normalised doses. For the contact-lens case, a first-order release profile was adopted as a simplified representation of sustained drug delivery through the corneal boundary. A time-dependent flux $J(t)$ was imposed on the inner side of the cornea:

$$J(t) = J_0 e^{-t/\tau} \tag{6}$$

Where $J_0$ is the initial flux value, $t$ is simulation time and $\tau$ is the release timescale, which has been normalised, so the conclusions remain independent of its specific value. The integral of $J(t)$ over release time must be equal to the initial dose $M_{dose}$ used across the four delivery strategies:

$$A_{lens} \int_0^{\infty} J(t)\, dt = M_{dose} \tag{7}$$

Where $A_{lens}$ is the inner corneal area. Eqn (10) yields a value of $J_0 = \frac{M_{dose}}{A_{lens}\tau}$.

### Saccadic movement

When implementing saccadic eye motion, the momentum equation was formulated in a rotating reference frame. This allowed the transient rotational terms induced by rapid angular motion to be incorporated directly into the momentum balance, namely the Coriolis, centrifugal, and Euler acceleration terms.

$$\rho\left(\frac{\partial \boldsymbol{u}}{\partial t} + (\boldsymbol{u}\cdot\nabla)\,\boldsymbol{u} + (2\boldsymbol{\omega}\times\boldsymbol{u}) + \big(\boldsymbol{\omega}\times(\boldsymbol{\omega}\times\boldsymbol{r})\big) + \dot{\boldsymbol{\omega}}\times\boldsymbol{r}\right) = -\nabla \mathrm{p} + \mu\nabla^2 u + \rho\boldsymbol{g}\left[1-\beta(T-T_{ref})\right] \quad (8)$$

Each saccade was modelled as a single rapid angular displacement of the eye from an initial primary gaze to a prescribed final gaze position and back to initial primary gaze. The initial orientation was defined as $\theta = 0$ at $t = 0$, corresponding to a primary gaze. The saccade amplitude denotes the final angular displacement of the eye relative to this initial orientation and the saccade duration was defined as the time taken to complete the imposed angular displacement. Angular displacement was modelled as a 5th-degree polynomial function of time[28]:

$$\theta(t) = c_0 + c_1 t + c_2 t^2 + c_3 t^3 + c_4 t^4 + c_5 t^5 \quad (9)$$

Where $c_0$ through $c_5$ are the polynomial coefficients. The initial conditions were defined as[29]:

$$\theta(0) = 0;\ \omega(0) = 0 \quad (10)$$

### Nanocarrier Transport

Nanocarrier transport was modelled using a Eulerian-Lagrangian approach. Nanocarriers were treated as discrete particles and their motion was governed by Newton's second law, while being advected by the computed aqueous humour velocity field. Under low $Re$ conditions, viscous drag dominates particle dynamics and is given by Stokes' law. Brownian motion was not included to account for nanoscale diffusion effects following the release into the aqueous humour domain due to a high Péclet number. Gravitational effects were incorporated through body-force terms acting on the particles.

## 3 Results and Discussion

The principal finding of this study is that aqueous humour hydrodynamics exert a significant and delivery-dependent influence on ocular drug transport beyond the effects of release kinetics alone. Even when the administered dose and characteristic delivery duration are normalised across delivery strategies, substantial differences remain in drug retention, spatial distribution and transport homogeneity. These findings demonstrate that the hydrodynamic environment of the anterior chamber constitutes an independent design consideration for ocular drug delivery systems.

### 3.1 Mesh Independence Study

A mesh independence study was conducted using four progressively refined mesh densities available automatically in the software: *coarse*, *normal*, *fine* and *finer*. Local mesh refinement was applied in regions expected to exhibit high spatial gradients, such as walls and porous domains (i.e., trabecular meshwork and implants).

Mesh convergence was assessed by comparing the average aqueous humour velocity magnitude within the anterior chamber and across the different mesh densities. Variations between the *fine* and *finer* meshes were below 2%, indicating grid-independent behaviour (Table 3). The fine mesh was therefore selected for all subsequent simulations as it provided sufficient numerical accuracy while maintaining reasonable computational cost. The differences found among the results obtained with these grid settings showed evidence of reaching the asymptotic range of convergence[30].

*Table 2 Grid independence check*

| | No. of elements | Average Velocity $\boldsymbol{u}$ ($\boldsymbol{\mu m/s}$) | Difference (%) |
|---|---|---|---|
| Coarse | 92099 | 35.621 | - |
| Normal | 243619 | 36.053 | 1.19% |
| Fine | 783189 | 36.373 | 0.88% |
| Finer | 1942121 | 36.625 | 0.68% |

### 3.2 Model verification and comparison with experimental observations

*In vivo* experimental validation within the eye's anterior chamber is particularly challenging because the area is small, optically complex, and physiologically delicate, thereby making direct measurements highly invasive and often impractical. Introducing probes or tracers can itself disturb the flow field, alter the intraocular pressure, or affect the drug transport dynamics, thereby compromising the very phenomena being measured. Furthermore, *in vivo* measurements are constrained by patient safety, ethical considerations, eye movements, blinking, and other factors. As a result, validation studies frequently rely on indirect measurements or comparisons with established numerical predictions. Qualitative and quantitative validation of the computed aqueous humour flow field was performed through comparison with previously reported numerical work. The predicted flow structures and velocity magnitudes were found to be in good agreement with published data.

Figure 2 presents the velocity magnitude contours and streamline distribution within the anterior segment with no saccade for both standing and supine positions, as well as the effect of saccadic movement on maximum corneal wall shear stress. Regarding velocity contours, the simulated flow exhibits a laminar regime with a low Reynolds number of approximately 0.5, driven by the pressure gradient between aqueous humour production at the ciliary body and drainage through the trabecular meshwork, and convective effects caused by the temperature difference between the cornea and the rest of the anterior segment's surface.

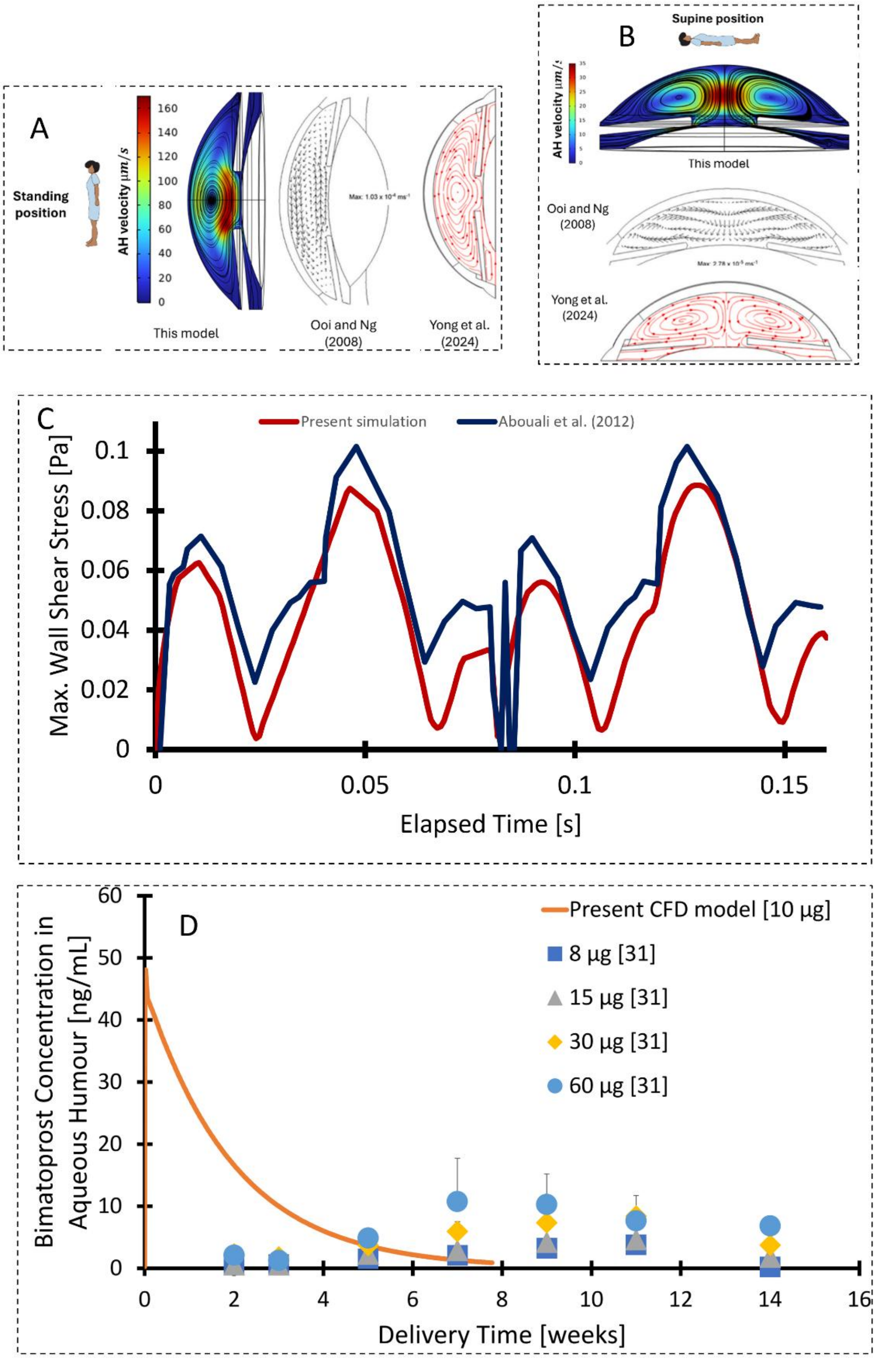

*Figure 2 Top left: static velocity profile and comparison with relevant literature. Top right: Idem for supine position. Middle graph: Comparison of maximum corneal wall shear stress between the present model and reported data. Comparison between the average bimatoprost concentration in the aqueous humour predicted by the present CFD model for a 10 μg intracameral implant and experimentally measured aqueous humour concentrations following intracameral administration of sustained-release bimatoprost implants containing 8, 15, 30 and 60 μg (adapted from [31])*

A localised region of elevated velocity was observed within the iridocorneal space immediately anterior to the pupil in the standing case, with a peak velocity of approximately 171 μm/s, similar in range to those found in the literature for studies incorporating physiological aqueous humour circulation and thermal buoyancy effects. This region forms the principal advective transport pathway within the anterior chamber and results from the combined effects of inflow conditions and thermal buoyancy. Velocity decreases substantially away from the inflow region, producing a predominantly viscous-dominated flow environment throughout the chamber.
Streamline analysis revealed a primary recirculating vortex occupying the anterior chamber in the standing position, with fluid motion directed from the inflow region towards the trabecular meshwork. For the supine position, a toroid-like ring vortex was observed. Other weak secondary recirculation zones were also observed. These flow structures closely resemble those reported in previous computational studies and play a critical role in governing the transport, mixing and clearance of both dissolved drug species and nanocarriers. The relatively low flow velocities predicted throughout the segment indicate a mixed advection-diffusion transport environment in which neither convection nor molecular diffusion can be neglected. These hydrodynamic conditions are expected to significantly influence the spatial distribution, retention and clearance of both dissolved drug species and nanocarriers, motivating the comparative transport analyses presented in the following sections. A case with a saccadic movement of 6° was run to obtain the maximum wall shear stress in the cornea. The results were compared satisfactorily with the literature.
In addition to the comparison of the simulated flow field with published numerical studies, the physiological plausibility of the predicted drug transport was assessed by comparing the simulated average bimatoprost concentration in the aqueous humour with published pharmacokinetic measurements following intracameral administration of sustained-release bimatoprost implants in dogs, using implants designed for zero-order release[31]. Figure 2 (bottom graph) shows that the concentrations predicted by the present CFD model for a 10 μg implant are of the same order of magnitude as those measured experimentally for implants containing 8–60 μg[31]. As caveats, experimental methodology to obtain data can explain differences in both experiments and simulations.

It should be noted that the present model employs an idealised first-order release profile to isolate the influence of aqueous humour hydrodynamics on drug transport, whereas the experimental biodegradable implant exhibits more complex release kinetics arising from polymer hydration, diffusion and biodegradation. Consequently, differences in the temporal evolution of the concentration profiles are expected and the comparison should not be interpreted as a quantitative pharmacokinetic validation. Rather, it provides an independent assessment that the simulated concentration magnitudes are physiologically plausible despite the simplified release model.

### 3.3 Hydrodynamics of diluted drug transport

#### Spatial distribution of diluted drug within the anterior chamber

Contour plots (Figure 3) reveal distinct spatial transport patterns associated with each administration route. The contact lens produces broadly distributed concentration fields throughout the anterior chamber. In contrast, implant-based delivery generates persistently localised concentration gradients near the release site. The injection case initially exhibits strong localisation immediately following administration but rapidly evolves towards a more uniform distribution as a consequence of aqueous humour circulation.

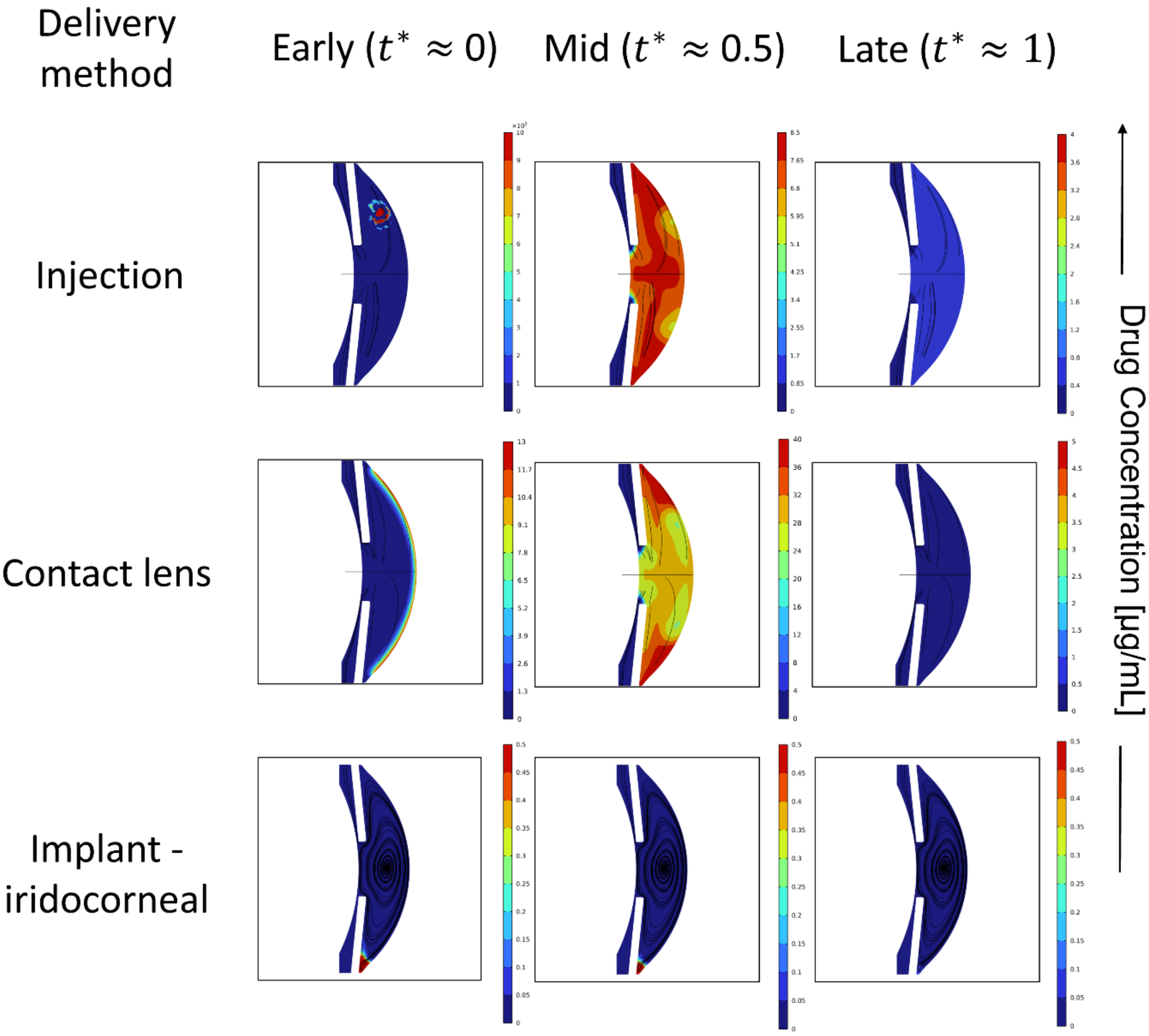

*Figure 3 Concentration contours for intracameral injection, contact-lens-mediated delivery, and implant-based delivery at representative normalised times ($t^* \approx 0; 0.5;$ and $1.0$). Time was normalised by the characteristic delivery duration associated with each administration route. A common colour scale was used for all panels to facilitate a direct comparison of spatial drug distributions. The contours illustrate the influence of administration route and source location on the evolution of intraocular concentration fields and complement the quantitative retention and spatial homogeneity analyses presented in Figures 4 and 5.*

## Spatial homogeneity of intraocular diluted drug delivery

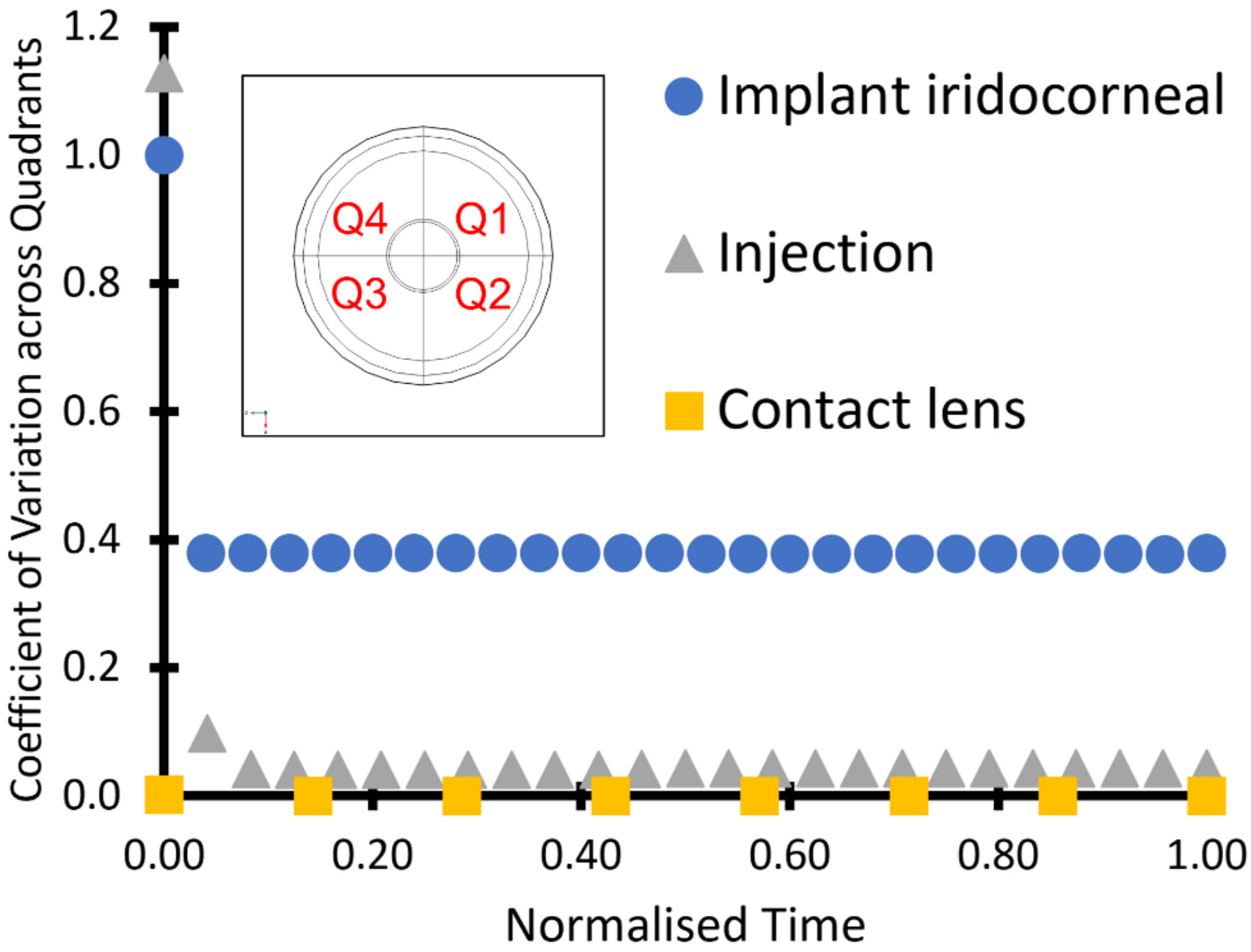

*Figure 4 Coefficient of variation (CV) of quadrant-averaged drug concentrations as a function of normalised time for intracameral injection, intracameral implantation in the iridocorneal angle and contact-lens delivery. The coefficient of variation was computed from the average concentrations in the four anterior-chamber quadrants of the eye ($CV = \sigma/\bar{c}$). Lower CV values indicate a spatially homogeneous drug distribution (i.e., little concentration differences between the eye's quadrants), whereas higher values correspond to greater concentration heterogeneity (i.e., all the drug localised in one or two quadrants of the anterior segment). The contact lens exhibits the most uniform distribution throughout the delivery period, while the implant configuration displays the greatest spatial variation. Intracameral injection produces an initially non-uniform distribution that rapidly approaches a nearly homogeneous state due to aqueous humour mixing.*

To quantify the spatial uniformity of drug distribution within the anterior chamber, a coefficient of variation (CV) of the quadrant-averaged concentrations was evaluated throughout the delivery period. Lower CV values correspond to more homogeneous concentration fields, whereas higher values indicate stronger spatial gradients and localised drug accumulation (Figure 4).

The contact lens exhibited the lowest CV values throughout the simulation, indicating a highly uniform drug distribution across the anterior chamber. This behaviour is expected given that drug entry occurs over the corneal boundary, promoting a distributed source that is rapidly mixed and cleared by the aqueous humour flow. In contrast, the implant displayed substantially higher CV values, reflecting the persistent concentration gradients generated by the localised release of drug near the trabecular meshwork.

The intracameral injection case exhibited an initially elevated CV, corresponding to the localised deposition of the drug immediately following administration. However, the coefficient of variation rapidly decreased and remained low thereafter, indicating efficient redistribution of the injected drug by aqueous humour circulation. This behaviour contrasts with the implant cases, where continuous localised release maintains concentration gradients throughout the delivery period. These results demonstrate that the mode of administration and the spatial location of the drug source play a dominant role in determining the uniformity of intraocular drug distribution.

**Ocular diluted drug retention within the aqueous humour**

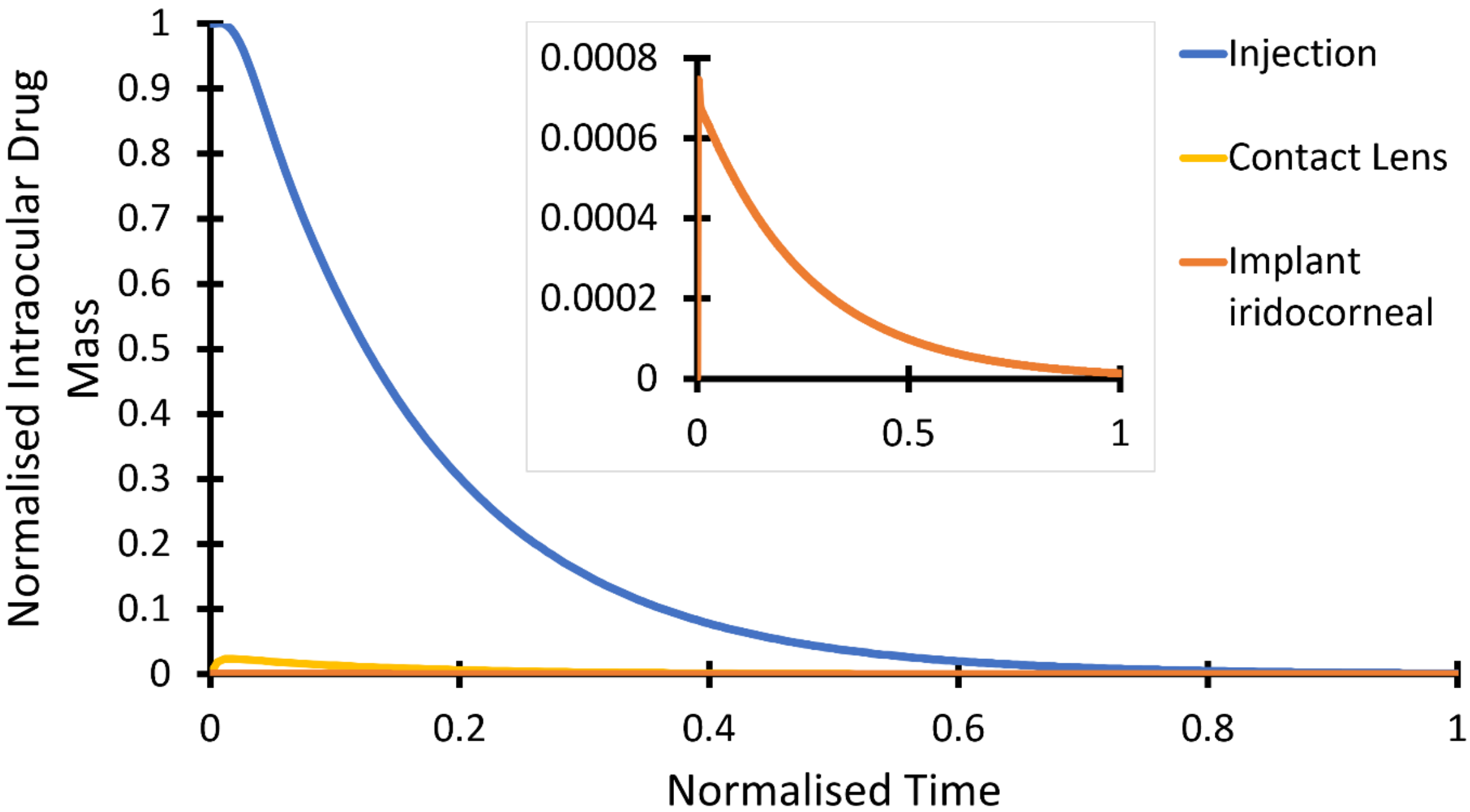

*Figure 5 Dimensionless intraocular drug retention profiles for intracameral injection, intracameral implantation and contact-lens delivery. Drug mass present within the aqueous humour was normalised by the total administered dose, while time was normalised by the characteristic delivery duration associated with each administration route. The persistence of distinct retention profiles following normalisation demonstrates that differences in intraocular drug fate cannot be attributed solely to delivery timescale. Instead, the results indicate that the mode and location of administration, together with their interaction with aqueous humour hydrodynamics, exert a significant influence on drug retention and clearance within the anterior chamber. The inset highlights the implant case, which exhibits substantially lower instantaneous intraocular drug masses owing to their sustained-release characteristics.*

To assess whether the observed differences in drug retention were primarily attributable to the imposed delivery method or to mixing within the anterior chamber, the drug mass present in the aqueous humour was normalised by the administered dose and plotted against time normalised by the characteristic delivery duration associated with each administration route (Figure 5). This procedure removes the influence of the absolute dose and delivery timescale and enables the direct comparison of the dimensionless retention behaviour of the different delivery strategies, in order to decide which method provides a better sustained release (i.e., flattest release curve).
Despite normalisation, substantial differences between the delivery routes remain evident. The injection case exhibits the highest initial normalised drug mass, followed by a gradual decay as the drug is removed by aqueous humour turnover. In contrast, the contact lens demonstrates an initial rise in intraocular drug mass before reaching an early maximum and subsequent decline, reflecting the competition between sustained delivery across the corneal boundary and hydrodynamic clearance. The implant case displays substantially lower normalised drug masses throughout the delivery period, with only a small fraction of the total administered dose being present within the aqueous humour at any given time.

The persistence of distinct retention profiles following normalisation indicates that the observed transport behaviour cannot be explained solely by differences in release duration. Instead, the results demonstrate that intraocular drug fate depends strongly on the mode and location of administration and its interaction with aqueous humour hydrodynamics. Consequently, while the characteristic delivery timescale determines the temporal distribution of the administered dose, aqueous humour flow remains a dominant factor governing the retention and clearance of drug within the anterior chamber.

The dimensionless retention profiles therefore support the hypothesis that aqueous humour hydrodynamics exert a significant influence on intraocular drug transport independently of the specific release timescale associated with each delivery strategy. Normalisation of both drug mass and delivery duration resulted in different retention curves for each administration route. This indicates that the observed differences cannot be solely attributed to variations in release timescale. Instead, the interaction between source location and aqueous humour hydrodynamics exerts a significant influence on intraocular drug retention, leading to differentiated transport and clearance behaviours even when equivalent doses are administered.

### Effect of isolated saccadic movements on diluted drug transport from iridocorneal implant

To assess the potential influence of transient eye motion, supplementary simulations incorporating a 50° saccadic eye movement were performed for the implant-based delivery configuration of diluted drug (Figure 8). Although the saccade generated observable perturbations in the aqueous humour streamlines, only minor changes in concentration distributions and molar transport rates were observed relative to the baseline case. For a large-amplitude saccade, the estimated Womersley number within the anterior chamber is of the order of $Wo = 10^1$, indicating that transient inertial effects are significant during the eye movement itself. However, the duration of the saccadic event is short compared with the characteristic timescales governing drug transport and clearance within the aqueous humour. Consequently, despite the transient modification of the flow field, the overall impact on long-term drug distribution was found to be negligible under the conditions investigated. The cumulative effects of repeated physiological eye movements over extended periods were not considered due to computational limitations.

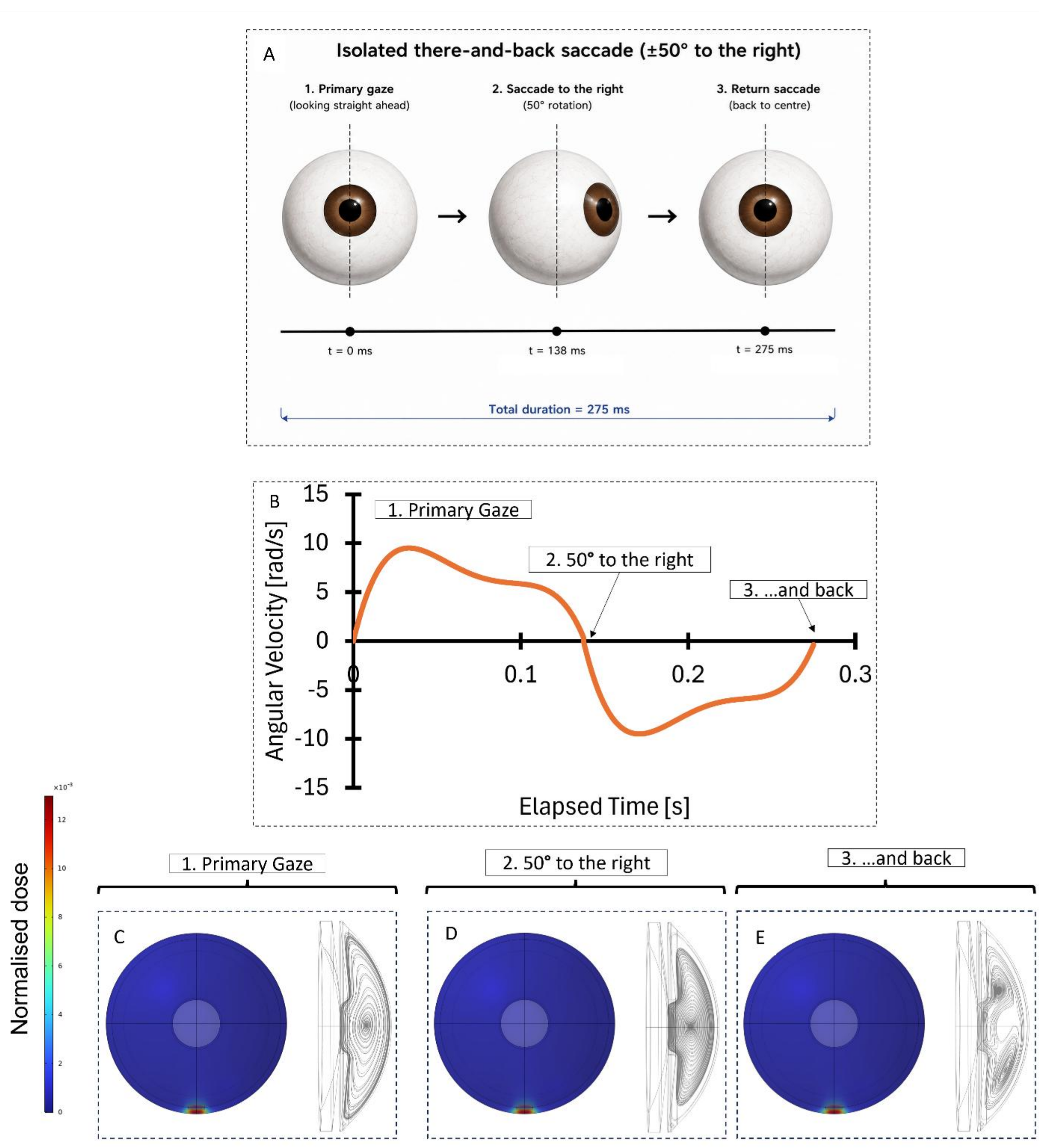

*Figure 6. The figure shows the angular velocity during an isolated there-and-back saccade (i.e., from straight centred gaze, rotation of 50° and back to straight centred gaze). Streamlines of aqueous humour flow and concentration maps have been represented, revealing changes in flow patterns which do not translate into diluted drug mixing. Concentration in µg/mL.*

### 3.4 Hydrodynamics of nanocarrier distribution

To characterise the dominant transport mechanisms governing nanoparticle motion, the particle Péclet number was estimated as $Pe = \frac{uL}{D_p}$, where $u$ and $L$ are characteristic aqueous humour velocity and length scales, respectively, and $D_p$ is the particle diffusivity. For a particle diffusivity of approximately $3\times10^{-12}$ $m^2/s$ and using representative anterior-chamber scales $u = 10^{-6}$ m/s, $L\sim 10^{-2}$ m, the resulting particle Péclet number is of order $O(10^4)$.

Since ($Pe \gg 1$), nanoparticle transport is dominated by advection associated with aqueous humour circulation, while Brownian diffusion plays a minor role. Consequently, particle trajectories primarily reflect the underlying intraocular flow field and the location of administration. The selected particle size of 200 nm therefore serves as a representative example of advection-dominated nanocarrier transport within the anterior chamber, rather than a specific formulation design to enable a comparison with diluted drug scenarios.

The trajectories of the nanocarriers provide qualitative insight into the mechanisms governing particle transport within the anterior chamber. In contrast to dissolved bimatoprost, which forms continuous concentration fields through the combined effects of advection and molecular diffusion, the nanocarriers predominantly follow the flow structures generated by aqueous humour circulation. This behaviour is consistent with the large particle Péclet number ($Pe\sim10^4$), indicating that nanoparticle transport is strongly advection-dominated and only weakly influenced by Brownian diffusion.

Distinct transport patterns are observed for the different administration routes (Figure 7). Following intracameral injection, particles are rapidly entrained by the circulating aqueous humour and distributed throughout a substantial portion of the anterior chamber. Contact-lens delivery generates a broadly distributed particle population entering from the corneal boundary, producing trajectories that extend throughout the chamber. In contrast, implant release exhibits more localised transport pathways associated with the vicinity of the target tissue. Therefore, particles released from the iridocorneal implant remain largely confined to the inferior region of the chamber.

The reduced diffusivity of the nanocarriers relative to dissolved bimatoprost results in a stronger dependence of transport behaviour on both aqueous humour hydrodynamics and source location. Consequently, the nanoparticle trajectories exhibit greater pathway dependence and reduced spatial spreading than the corresponding dissolved-drug simulations. These observations suggest that administration route may play a more important role in determining intraocular nanoparticle distribution than in the case of dissolved therapeutic agents. The qualitative transport patterns observed in Figure 6 are further quantified through quadrant averaged spatial homogeneity analysis in the next section.

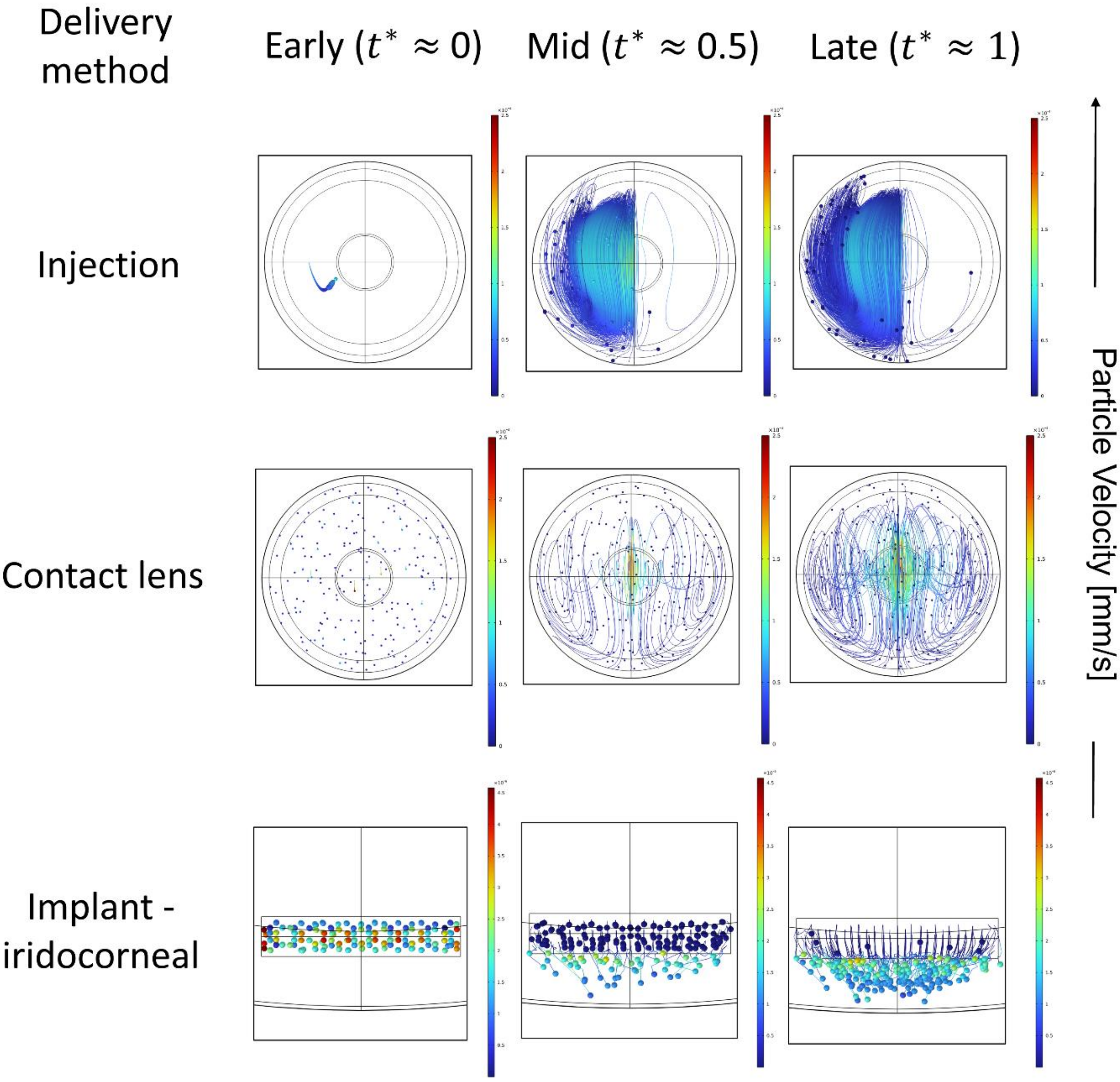

*Figure 7 Representative trajectories of nanocarriers delivered by intracameral injection, contact-lens administration, and implant placement in the iridocorneal angle. Particle trajectories are shown at early, intermediate and late dimensionless times and are coloured according to particle velocity magnitude. The trajectories illustrate the transport pathways established by aqueous humour circulation and reveal distinct patterns of nanoparticle dispersion associated with each delivery strategy.*

**Spatial homogeneity of intracameral nanocarrier delivery**

Figure 8 demonstrates a substantial increase in spatial heterogeneity for nanoparticle transport relative to the dissolved-drug simulations. Although the ranking of the delivery strategies remains unchanged, with contact-lens administration producing the most uniform distributions (CV → 0) and implant-based delivery producing the least uniform distributions (CV ≥ 1), the coefficient of variation is larger for all nanoparticle cases with respect to the diffusion dominated counterparts.

This behaviour is consistent with the reduced diffusivity of the nanocarriers. The corresponding particle Péclet number is of order (10^4), indicating strongly advection-dominated transport.

Consequently, particles remain preferentially confined to the flow pathways established by the aqueous humour circulation, while transport between quadrants is substantially reduced. In contrast, dissolved bimatoprost experiences greater diffusive spreading, which acts to smooth concentration gradients and promote spatial homogenisation.

The largest coefficients of variation are observed for the implant configuration. Intracameral injection produces intermediate levels of heterogeneity, reflecting rapid entrainment of particles by the aqueous humour flow, but limited subsequent mixing between quadrants. Contact-lens delivery remains the most spatially uniform strategy owing to the distributed nature of the corneal source; however, the increase in the coefficient of variation relative to the dissolved-drug simulations indicates that reduced particle diffusivity still limits complete homogenisation.

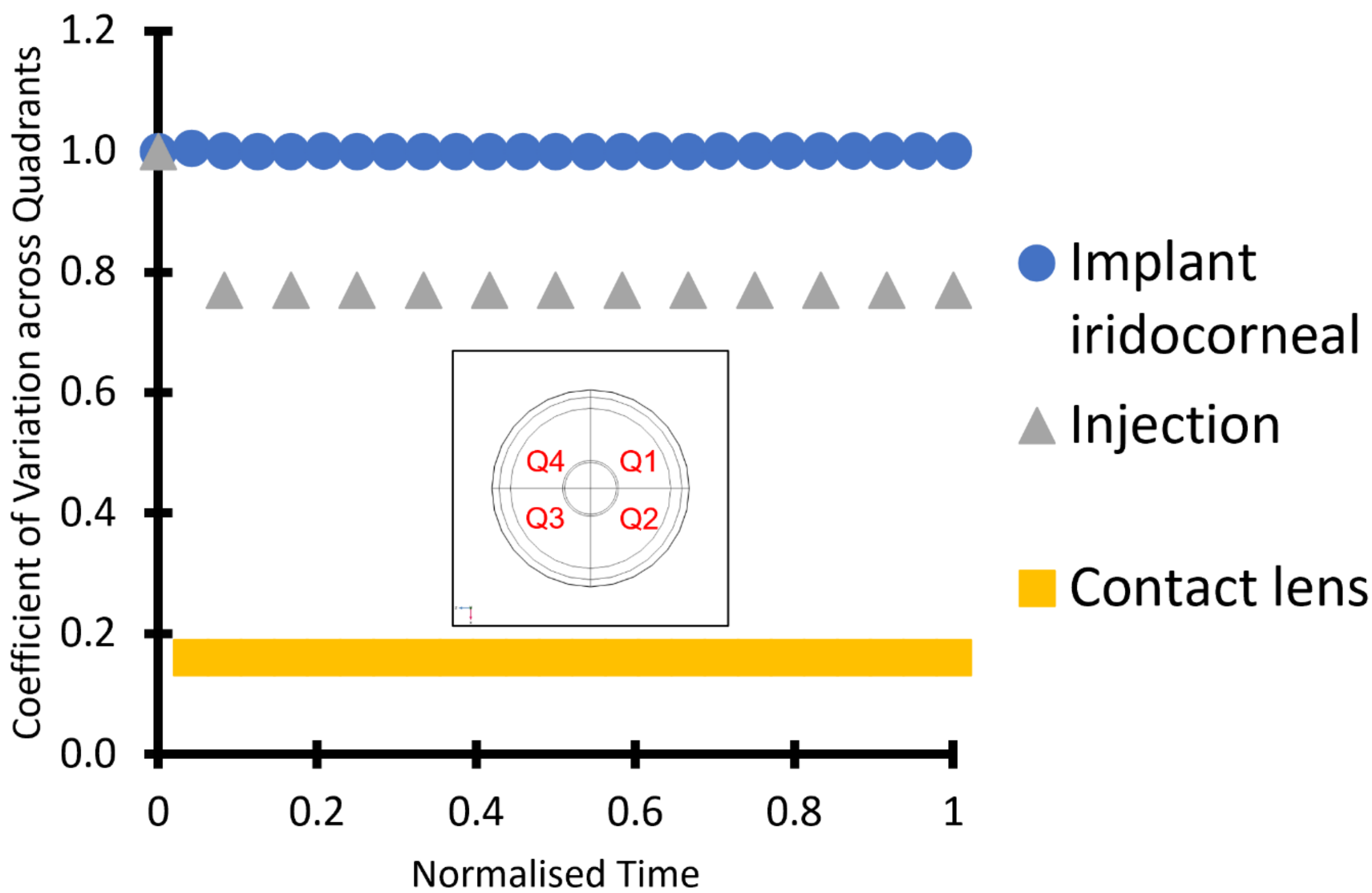

*Figure 8: Coefficient of variation (CV) of quadrant-averaged nanoparticle concentrations as a function of normalised time for intracameral injection, contact-lens delivery, and implant placement in the iridocorneal angle. The coefficient of variation was computed from the average nanoparticle concentrations in the four anterior-chamber quadrants ($CV = \sigma/\bar{c}$), where larger values indicate greater spatial heterogeneity. Time was normalised by the characteristic delivery duration associated with each administration route. The contact lens exhibits the most homogeneous nanoparticle distribution throughout the delivery period, whereas implant delivery maintains substantial spatial heterogeneity owing to localised release site. Intracameral injection produces intermediate levels of heterogeneity, reflecting rapid particle entrainment by the aqueous humour flow coupled with limited inter-quadrant mixing.*

## 4. Conclusions

A CFD framework was developed to investigate the influence of aqueous humour hydrodynamics on the transport of dissolved drug and nanoparticle-based formulations delivered through intracameral injection, intracameral implantation and contact-lens administration for glaucoma therapy. Predicted aqueous humour concentrations were found to be consistent with published experimental measurements, providing additional confidence in the physiological realism of the proposed CFD framework. By comparing multiple delivery strategies within a common anatomical and hydrodynamic environment, this study provides insight into the role of administration route, source location and transport mechanism in determining the intraocular drug fate.

The simulations demonstrate that aqueous humour circulation exerts a significant influence on both the spatial distribution and retention of therapeutic agents within the anterior chamber. Concentration contour maps and quantitative homogeneity metrics revealed substantial differences between delivery strategies. Contact-lens delivery produced the most spatially uniform drug distributions, whereas implant-based delivery generated persistent concentration gradients associated with the localised nature of the release source. Intracameral injection exhibited intermediate behaviour, with rapid redistribution of the administered drug by the aqueous humour flow.

Dimensionless analysis of intraocular drug retention showed that distinct retention behaviours persist even after normalisation with respect to administered dose and characteristic delivery duration. This indicates that differences between delivery strategies cannot be attributed solely to the timescale of drug administration. Instead, the interaction between source location and aqueous humour hydrodynamics plays a fundamental role in governing intraocular drug transport and clearance.

The nanoparticle simulations revealed a substantially stronger dependence on hydrodynamic transport than the corresponding dissolved-drug cases. Owing to their reduced diffusivity, the 200 nm nanocarriers exhibited large particle Péclet numbers, resulting in strongly advection-dominated transport. Consequently, nanoparticle motion remained closely associated with the underlying flow pathways, leading to greater spatial heterogeneity and more persistent localisation effects than those observed for dissolved bimatoprost. Although the relative ranking of the delivery strategies remained unchanged, the magnitude of the differences increased considerably for nanoparticle transport.

Overall, this work demonstrates that aqueous humour hydrodynamics are not simply a passive transport mechanism but an active determinant of ocular drug delivery performance. Consequently, hydrodynamic transport should be considered alongside formulation design and release kinetics when developing sustained ophthalmic drug delivery systems.

# References

1. Peng, Y. *et al.* Global, regional and national burden of glaucoma from 1990 to 2021 and projections to 2050: a retrospective cross-sectional study. *BMJ Open* **16**, e108975 (2026).
2. Belamkar, A. *et al.* Sustained release glaucoma therapies: Novel modalities for overcoming key treatment barriers associated with topical medications. *Annals of Medicine* vol. 54 343–358 Preprint at https://doi.org/10.1080/07853890.2021.1955146 (2022).
3. Li, C. C. & Chauhan, A. Modeling ophthalmic drug delivery by soaked contact lenses. *Ind. Eng. Chem. Res.* **45**, 3718–3734 (2006).
4. Li, S., Chen, L. & Fu, Y. Nanotechnology-based ocular drug delivery systems: recent advances and future prospects. *Journal of Nanobiotechnology* vol. 21 Preprint at https://doi.org/10.1186/s12951-023-01992-2 (2023).
5. Agrahari, V. *et al.* A comprehensive insight on ocular pharmacokinetics. *Drug Delivery and Translational Research* vol. 6 735–754 Preprint at https://doi.org/10.1007/s13346-016-0339-2 (2016).
6. Jani, H. S. *et al.* An Update on Novel Drug Delivery Systems for the Management of Glaucoma. *Pharmaceutics* vol. 17 Preprint at https://doi.org/10.3390/pharmaceutics17081087 (2025).
7. *Current Management of Glaucoma and the Need for Complete Therapy*. www.ajmc.com.
8. ’Deitz, G. C. Y. C. E. ’. Sustained-delivery Glaucoma Treatments. *GLAUCOMA PHYSICIAN* 14–16 (2024).
9. Paleel, F., Qin, M., Tagalakis, A. D., Yu-Wai-Man, C. & Lamprou, D. A. Manufacturing and characterisation of 3D-printed sustained-release Timolol implants for glaucoma treatment. *Drug Deliv. Transl. Res.* **15**, 242–252 (2025).
10. Qin, M. *et al.* Precise Control of Drug Release in Machine Learning-Designed Antibody-Eluting Implants for Postoperative Scarring Inhibition in Glaucoma. *Adv. Healthc. Mater.* https://doi.org/10.1002/adhm.202502689 (2026) doi:10.1002/adhm.202502689.
11. Gaudana, R., Ananthula, H. K., Parenky, A. & Mitra, A. K. Ocular drug delivery. *The AAPS journal* vol. 12 348–360 Preprint at https://doi.org/10.1208/s12248-010-9183-3 (2010).
12. Desai, A. R. *et al.* Co-delivery of timolol and hyaluronic acid from semi-circular ring-implanted contact lenses for the treatment of glaucoma:: In vitro and in vivo evaluation. *Biomater. Sci.* **6**, 1580–1591 (2018).
13. El Shaer, A. *et al.* Nanoparticle-laden contact lens for controlled ocular delivery of prednisolone: Formulation optimization using statistical experimental design. *Pharmaceutics* **8**, (2016).
14. Yang, C. *et al.* Nanoparticles in ocular applications and their potential toxicity. *Frontiers in Molecular Biosciences* vol. 9 Preprint at https://doi.org/10.3389/fmolb.2022.931759 (2022).
15. Heys, J. J. & Barocas, V. H. A Boussinesq model of natural convection in the human eye and the formation of Krukenberg’s spindle. *Ann. Biomed. Eng.* **30**, 392–401 (2002).

16. Villamarin, A. *et al.* 3D simulation of the aqueous flow in the human eye. *Med. Eng. Phys.* **34**, 1462–1470 (2012).

17. Martínez Sánchez, G. J., Escobar del Pozo, C., Rocha Medina, J. A., Naude, J. & Brambila Solorzano, A. Numerical simulation of the aqueous humor flow in the eye drainage system; a healthy and pathological condition comparison. *Med. Eng. Phys.* **83**, 82–92 (2020).

18. Karimi, A. *et al.* Modeling the biomechanics of the conventional aqueous outflow pathway microstructure in the human eye. *Comput. Methods Programs Biomed.* **221**, (2022).

19. Sebastia-Saez, D., Luo, J., Qin, M., Chen, T. & Yu-Wai-Man, C. Advancing glaucoma research with multiphysics continuum mechanics modelling: Opportunities and open challenges. *Journal of the Mechanical Behavior of Biomedical Materials* vol. 173 Preprint at https://doi.org/10.1016/j.jmbbm.2025.107240 (2026).

20. Wang, W., Qian, X., Song, H., Zhang, M. & Liu, Z. Fluid and structure coupling analysis of the interaction between aqueous humor and iris. *Biomed. Eng. Online* **15**, (2016).

21. Ferreira, J. A., De Oliveira, P., Da Silva, P. M. & Murta, J. N. Numerical simulation of aqueous humor flow: From healthy to pathologic situations. *Appl. Math. Comput.* **226**, 777–792 (2014).

22. Fernández-Vigo, J. I. *et al.* Computational simulation of aqueous humour dynamics in the presence of a posterior-chamber versus iris-fixed phakic intraocular lens. *PLoS One* **13**, (2018).

23. Cui, L. *et al.* Study on the correlation between iris blood flow, iris thickness and pupil diameter in the resting state and after pharmacological mydriasis in patients with diabetes mellitus. *BMC Ophthalmol.* **24**, (2024).

24. Fda & Cder. *HIGHLIGHTS OF PRESCRIBING INFORMATION These Highlights Do Not Include All the Information Needed to Use DURYSTA TM Safely and Effectively. See Full Prescribing Information for DURYSTA TM. DURYSTA TM (Bimatoprost Implant), for Intracameral Administration*. www.fda.gov/medwatch.

25. Pan, Z. & Ding, J. Poly(lactide-co-glycolide) porous scaffolds for tissue engineering and regenerative medicine. *Interface Focus* vol. 2 366–377 Preprint at https://doi.org/10.1098/rsfs.2011.0123 (2012).

26. Abu-Hassan, D. W., Acott, T. S. & Kelley, M. J. *The Trabecular Meshwork: A Basic Review of Form and Function*.

27. Cai, J. C., Chen, Y. L., Cao, Y. H., Babenko, A. & Chen, X. Numerical study of aqueous humor flow and iris deformation with pupillary block and the efficacy of laser peripheral iridotomy. *Clinical Biomechanics* **92**, (2022).

28. Abouali, O., Modareszadeh, A., Ghaffarieh, A. & Tu, J. Investigation of saccadic eye movement effects on the fluid dynamic in the anterior chamber. *J. Biomech. Eng.* **134**, (2012).

29. Repetto, R., Stocchino, A. & Cafferata, C. Experimental investigation of vitreous humour motion within a human eye model. *Phys. Med. Biol.* **50**, 4729–4743 (2005).

30. Richardson, L. The approximate arithmetical solution by finite differences of physical problems involving differential equations, with an application to the stresses in a masonry dam. *Phylosophical transactions of the Royal Society A* **210**, 307–357 (1911).

31. Shen, J., Robinson, M. R., Struble, C. & Attar, M. Nonclinical pharmacokinetic and pharmacodynamic assessment of bimatoprost following a single intracameral injection of sustained-release implants. *Transl. Vis. Sci. Technol.* **9**, 1–11 (2020).